# Optimization and Dynamic Analysis of a Vibro-Impact Nonlinear Energy Sink with Electromagnetic Coil for Vibration Suppression and Energy Harvesting

Ali Abdollahi

Tarbiat Modares University Tehran P.O. Box 14115-177, Iran

**Abstract**

This study investigates a system comprising a linear oscillator (LO) equipped with a Vibro-Impact Nonlinear Energy Sink (VI-NES) and a coil. The LO's damping and stiffness coefficients are represented by ( C ) and ( k ), respectively, while a ball inside the LO moves within a cavity, colliding with the walls at both ends. The primary focus is on optimizing the cavity length (( L_c )) and the coefficient of restitution (( \kappa )) and coil parameter () using a Genetic Algorithm (GA). The motion equations for the LO incorporating the VI-NES are derived, considering the electromagnetic force (( F_e )) and contact force (( F_c )). The study also explores the energy harvesting circuit, which generates electric power by connecting a load resistance to the coil, transforming mechanical energy into electrical energy. The dynamic behavior of the system is analyzed, highlighting the transition from predictable patterns to chaotic responses. Efficiency metrics are defined to measure the VI-NES's performance in absorbing and dissipating excitation energy. The optimization formulation addresses the influence of initial conditions, proposing a multi-objective approach to handle uncertainties. Validation against existing studies confirms the reliability of the proposed optimization method. The results demonstrate that optimizing both ( L_c ) and ( \kappa ) using GA effectively minimizes the amplitude response of the LO, outperforming approximate methods.

E-mail addresse: a.abdolahi@modares.ac.ir (A. Abdollahi)

# 1. Introduction

Various types of absorbers are employed to reduce system vibrations. Some of these absorbers are designed to not only mitigate vibrations but also to capture energy from them, which is particularly advantageous in scenarios where energy efficiency is crucial [1, 2]. Recently, there has been significant research into new methods of Targeted Energy Transfer (TET) using passive controllers for both energy harvesting and vibration control [3-8]. TET can be achieved through different methods, such as the classical linear tuned mass damper (TMD) approach, which involves adding a small mass, a damper, and a spring to the primary system to control its vibrations [9]. Consequently, the vibration energy from the primary system is transferred to the secondary system, where it is dissipated through damping.

Numerous studies on Targeted Energy Transfer (TET) involve a system where a linear oscillator is coupled with a Nonlinear Energy Sink (NES) [10-15]. NESs are characterized by their nonlinear stiffness and damping properties, enabling them to adapt to a broad spectrum of vibration frequencies. This adaptability makes them highly efficient at absorbing energy from various vibrations. The initial demonstration of TET involved coupling a nonlinear oscillator (without a linear component) to a linear system [16]. It was shown that under certain conditions, a nonlinear system could absorb energy from a linear structure subjected to free vibrations (due to initial velocity). However, this study did not address a nonlinear oscillator with a linear term, which was later explored in [17]. In that research, it was demonstrated that a nonlinear energy sink (NES) incorporating both linear and nonlinear elements could also achieve TET by appropriately selecting the spring coefficient.

Vibro-impact (VI) is a device that

typically includes components that experience impacts, such as a mass striking a stop. These impacts facilitate the rapid transfer of energy from the primary structure to the energy sink, thereby enhancing the overall damping effect. They are a type of nonlinear energy sink (NES) used for targeted energy transfer (TET) in vibration control and energy harvesting applications. VI systems feature a ball that moves freely within a cavity and collides with the walls, creating a mechanism for energy dissipation and transfer. One advantage of these systems is their simplicity compared to NES and TMD systems, as they do not require springs or dampers. However, VI systems present analytical challenges due to their non-smooth and discontinuous dynamics. Most existing studies on VI-NES utilize the method of multiple scales [18, 19], which is applicable only when the mass of the ball is significantly smaller than the mass of the main system [20]. This method helps determine the optimal conditions for TET, such as the cavity length and the number of impacts per period [21].

A vibro-impact system with an electromagnetic coil is a cutting-edge device engineered for both vibration control and energy harvesting [22, 23]. It integrates the principles of vibro-impact dynamics with electromagnetic transduction to efficiently dissipate and convert energy. The system features a mass linked to an electromagnetic coil, which transforms the kinetic energy from impacts into electrical energy [24]. This dual-purpose design enables the system to suppress vibrations while simultaneously harvesting energy. The synergy of vibro-impact and electromagnetic mechanisms introduces nonlinear dynamics, allowing the system to adapt to a broad range of vibration frequencies, making it highly effective in absorbing energy from various vibrations.

The performance of an absorber can be greatly affected by uncertainties in its parameters. Consequently, relying on a deterministic optimization method can result in an unstable optimal condition, where these uncertainties significantly diminish the effectiveness of Targeted Energy Transfer (TET) [25]. Therefore, it is essential to adopt a design approach that determines the optimal values for parameters such as the restitution coefficient and cavity length, while accounting for uncertainties in both design and random parameters. The literature indicates that there has been limited research in this area. Missoum and Boroson [25] introduced a probabilistic design method for a nonlinear energy sink (NES) that considers random design and aleatory variables. Their method involves identifying regions within the design and aleatory space that correspond to different levels of NES efficiency. Pidaparthi and Missoum [26] investigated the reduction of limit cycle oscillations (LCOs) in a two-degree-of-freedom airfoil using an optimally designed NES, exploring several stochastic optimization problems aimed at maximizing the mean reduction of LCOs. Qian and Chen [27] developed an analytical method to optimize displacement reduction in a linear oscillator by incorporating an impact element and an inerter. Wu et al. examined the power spectral characteristics of a structure's vibration with an NES in the frequency domain [28]. Bao et al. utilized the mass amplification characteristic of an inerter to design an inertial nonlinear energy sink (INES), which is used to mitigate the random response of a spring-mass-damper system under random excitation [29]. Despite these studies, a method to minimize the system's energy using a VI-NES under random excitation is still lacking.

The main contribution of this paper is the development of a stochastic optimization approach for designing key parameters of a vibration absorber, including coil parameters, restitution coefficient, and cavity length. This method enables the design of absorbers that are effective across a wide range of initial conditions, in contrast to traditional deterministic optimization techniques, which are limited to specific initial velocities. By leveraging stochastic optimization, the proposed approach enhances the adaptability and robustness of vibration absorbers in practical applications.

The first section of the paper presents a mathematical model of the system, laying the groundwork for understanding its dynamics. Following this, the dynamic regime of the optimal system is introduced, outlining the key behaviors and characteristics that define its performance. In the subsequent section, the optimization method is formulated, detailing the approach taken to design the vibration absorber's key parameters. The results from both deterministic and stochastic optimization methods are then compared, highlighting the advantages of the latter in achieving effectiveness across a broader range of initial conditions. The methodology is further validated by referencing previous works, demonstrating its robustness and reliability. Finally, the paper concludes with a summary of findings and implications for future research in vibration absorber design.

## 2. Model description

The system consists of a linear oscillator (LO) equipped with a VI-NES. The LO's damping and stiffness coefficient constant are represented by C and k, respectively. Inside the LO, a ball moves within a cavity, colliding with the walls at both ends. The length of the cavity and the coefficient of restitution are denoted by $L_c$ and $\kappa$, respectively. These parameters, $L_c$ and $\kappa$, are the primary

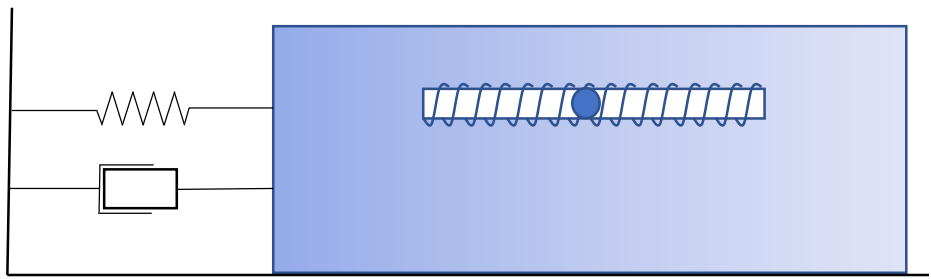

Fig1 Schematic of an LO with a VI-NES and coil focus of this study's design. The masses of the LO and the ball are indicated by M and m, respectively. The friction between the cavity and the impact element is negligible. The motion equation for the LO incorporating a VI-NES (as shown in Fig. 1) is given by:

$$M\frac{d^2}{dt^2}x_1(t) + c\frac{d}{dt}x_1(t) + kx_1(t) = F_e + \mathcal{F}_c \quad (1)$$
$$m\frac{d^2}{dt^2}x_2(t) + \mathcal{F}_c + F_e = 0$$

where $F_e$ represents the electromagnetic force generated by the interaction between the oscillating magnet and the coil, and $\mathcal{F}_c$ denotes the contact force between the magnet and the gap walls.

The harvesting circuit shown in Fig. 2 generates electric power by connecting a load resistance, $R_{load}$ to the coil, allowing a current (i) to flow. This current induces the electromagnetic force $F_e$ described in Eq. (1), which counteracts the relative motion between the magnet and the linear oscillator. By opposing $F_e$, mechanical energy is transformed into electrical energy. According to Lenz's law, $F_e$ is calculated as follows:

$$F_e = k_t i \quad (2)$$

where $k_t$ is referred to as the transduction factor or flux linkage gradient. For simplicity, it is assumed that the magnetic field generated by the magnet does not vary with time, making $k_t$ a constant. Additionally, the current (i) in the coil is determined as follows:

$$i = \frac{U_s}{R_{load} + R_{coil}} \quad (3)$$

where $U_s$ represents the generated voltage or induced electromotive force. According to Faraday's law, this can be expressed as the product of the flux linkage gradient $k_t$ and the relative velocity:

$$U_s = k_t(\frac{d}{dt}x_2(t) - \frac{d}{dt}x_1(t)) \quad (4)$$

The combination of equations (3) to (5) yields:

$$F_e = \mathbb{C}_e(\frac{d}{dt}x_1(t) - \frac{d}{dt}x_2(t)), \quad \mathbb{C}_e = \frac{k_t^2}{R_{load}+R_{coil}} \tag{5}$$

By substituting Eq(6) to Eq(1) one can obtain:

$$M\frac{d^2}{dt^2}x_1(t) + c\frac{d}{dt}x_1(t) + kx_1(t) + \mathbb{C}_e(\frac{d}{dt}x_1(t) - \frac{d}{dt}x_2(t)) = \mathcal{F}_c \tag{6}$$
$$m\frac{d^2}{dt^2}x_2(t) - \mathbb{C}_e(\frac{d}{dt}x_1(t) - \frac{d}{dt}x_2(t)) + \mathcal{F}_c = 0$$

$$\frac{d^2}{d\tau^2}x_1(\tau) + \varepsilon\xi\frac{d}{d\tau}x_1(\tau) + x_1(\tau) + \mathsf{C}_e(\frac{d}{d\tau}x_1(\tau) - \frac{d}{d\tau}x_2(\tau)) = F_c \tag{7}$$
$$\varepsilon\frac{d^2}{d\tau^2}x_2(\tau) - \mathsf{C}_e(\frac{d}{d\tau}x_1(\tau) - \frac{d}{d\tau}x_2(\tau)) + F_c = 0$$

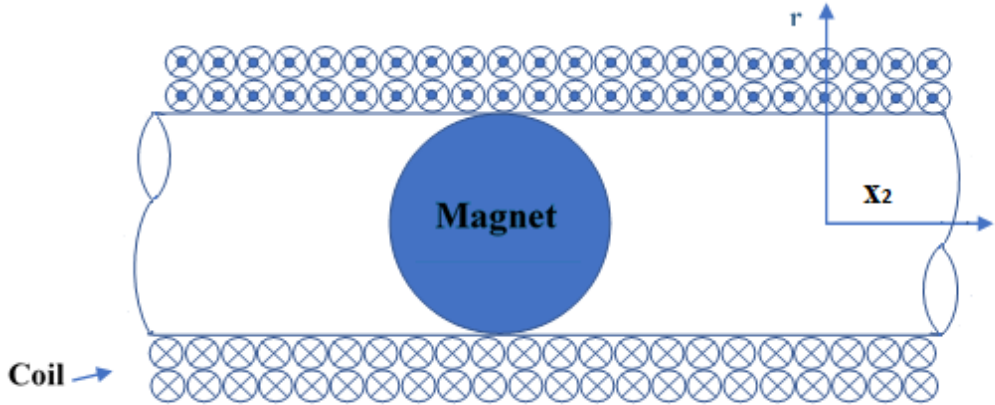

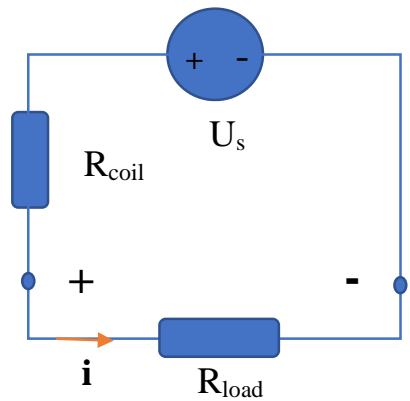

Fig. 2 The energy harvesting circuit.

Where:

$$\varepsilon = \frac{m}{M}, \omega^2 = \frac{k}{M}, \varepsilon\xi = \frac{C}{M\omega}, F_c = \frac{\mathcal{F}_c}{M\omega^2}, \mathbb{C}_e = \frac{\mathcal{C}_e}{M\omega}, t = \omega\tau \tag{8}$$

When $|x_1 - x_2| = L_c$, impacts take place. The post-impact state of the LO and the impact element is determined by applying the principle of total momentum conservation for $|x_1 - x_2| = L_c$:

$$x_1^+ = x_1^-, x_1^+ = x_1^-, \frac{d}{dt}x_1^+(t) - \frac{d}{dt}x_2^+(t) = -\kappa(\frac{d}{dt}x_1^-(t) - \frac{d}{dt}x_2^-(t)) \tag{9}$$

The symbols (+) and (-) denote the moments immediately after and before the impact, respectively. New variables are introduced to represent the displacement of the center of mass and the relative displacement between the VI-NES and the LO, as follows:

$$X = x_1 + \varepsilon x_2, w = x_1 - x_2 \tag{10}$$

By inserting Eq. (3) and (4) into Eq. (1), Eq. (5) is derived:

$$\begin{cases} X_{\tau\tau} + \varepsilon\lambda(X_\tau + \varepsilon w_\tau) + \frac{(X + \varepsilon w)}{1 + \varepsilon} = 0 \\ w_{\tau\tau} + \varepsilon\lambda(X_\tau + \varepsilon w_\tau) + \frac{(X + \varepsilon w)}{1 + \varepsilon} + c_e w_\tau = f_c \end{cases} \tag{11}$$

Where: \hfill (12)

$$\varpi = \frac{1}{1 + \varepsilon}, \lambda = \frac{\xi}{\varpi}, c_e = \frac{(1 + \varepsilon)^{\frac{3}{2}}}{\varepsilon}\mathbb{C}_e, f_c = \frac{(1 + \varepsilon)^2}{\varepsilon}F_c$$

## 3. Dynamical analysis of the system

The plot illustrates the dynamic behavior of a vibro-impact system, drawing on findings from reference [17]. Initially, the system exhibits a predictable pattern characterized by two distinct impacts per cycle at the motion onset (It is a sign of optimal response). This behavior transitions into a chaotic response as the motion progresses, a phenomenon vividly captured in Fig. 3, which presents the wavelet transform of the relative displacement between the magnet and the primary oscillator.

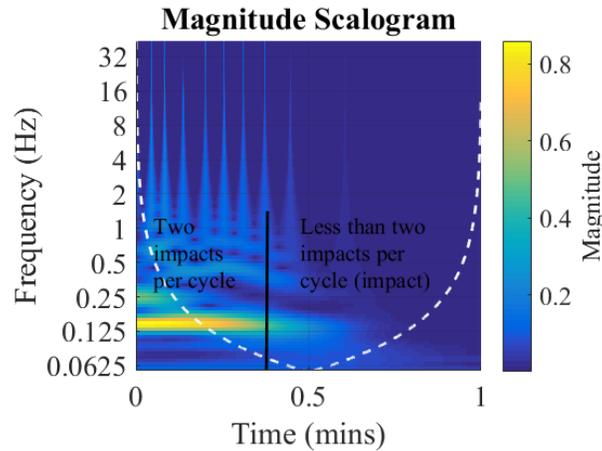

Fig. 3 Wavelet analysis of the optimal relative movement between a linear oscillator and a magnet. $c_e$=0.05, e=0.05, c = 0.2, $L_c$ = 0.99, $\kappa$ = 0.54, $\dot{x}_{1,0}$ =0.5

In Fig. 4a, the relative displacement is further detailed, highlighting the interactions between the magnet and oscillator throughout the motion. The subsequent Fig. 4b expands on this by illustrating just over 5 seconds of motion, during which three impacts occur, marking a shift in the system's dynamics.

Fig. 4c provides critical insights into the velocities of both components post-impact. It reveals that following the first impact, the velocity of the linear oscillator decreases, while the velocity of the impact element—the magnet—experiences a significant increase. This interplay is important for understanding energy transfer within the system. As the magnet moves within the cavity, its velocity gradually diminishes due to the influence of the coil effect. Notably, if the coil were absent, the magnet's velocity would remain constant, resulting in a horizontal line segment between the first and second impacts, indicating no change in velocity.

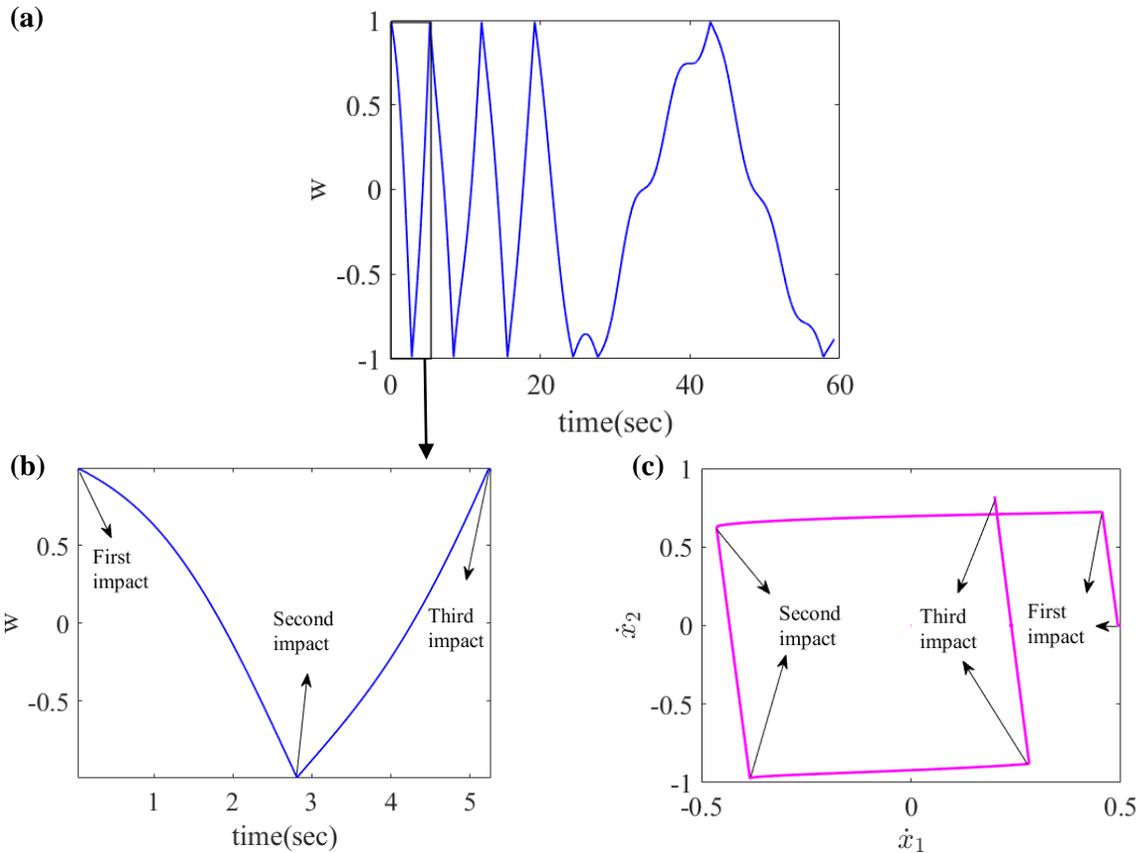

Fig. 4 (a) The optimal relative movement between a linear oscillator and a magnet. (b) Zoomed-in view of the Fig. 4a (c) Graph depicting the varying velocities of the linear oscillator and the magnet.

These figures collectively illustrate a complex interplay of forces and motions within a vibro-impact system, showcasing both predictable patterns and chaotic behavior as influenced by the presence of coils and the nature of impacts.

Fig. 5a illustrates the time history of the system over 60 seconds, demonstrating the absorber's effectiveness in suppressing the vibration of the linear oscillator. Fig. 5b presents a 3D plot depicting the velocities of the magnet and the linear oscillator over time. When viewed along the time axis (Fig. 5c), the figure reveals the chaotic response of the system.

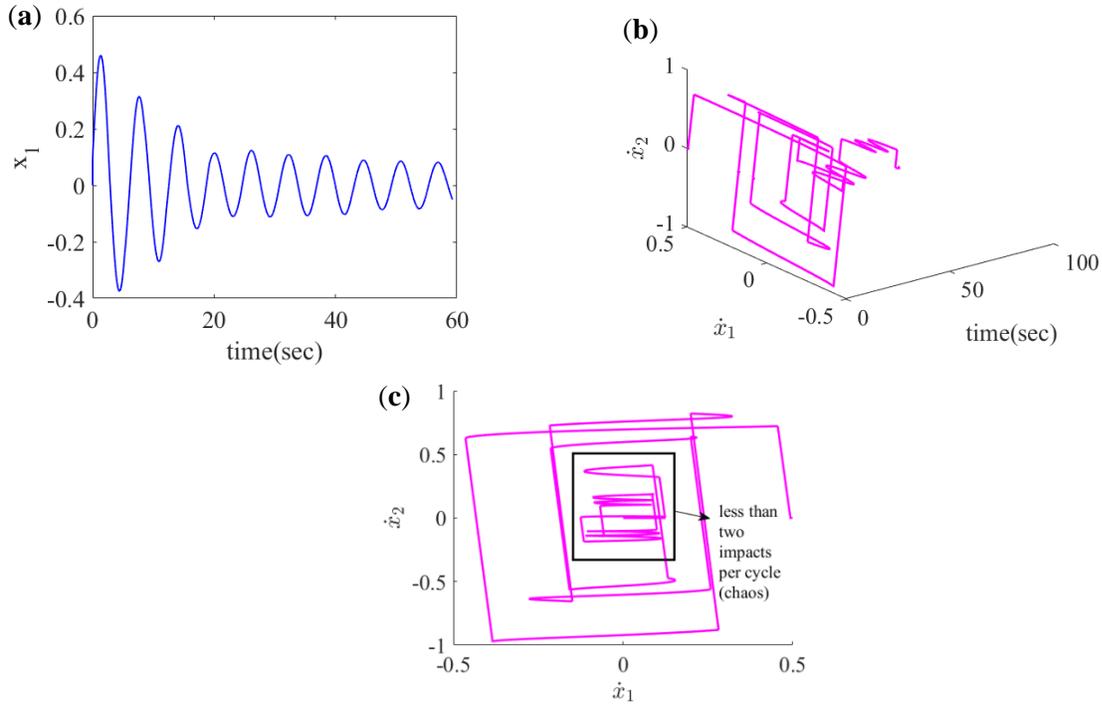

Fig. 5 (a) Time history (b) The varying velocities of the linear oscillator and the magnet over time (c) The varying velocities of the linear oscillator and the magnet

Fig. 6 displays the phase trajectory of the VI-NES over 30 seconds. The plot highlights the initial motion (solid line) and the final motion (dashed line). A noticeable decrease in the system's relative velocity, and consequently its kinetic energy, is evident.

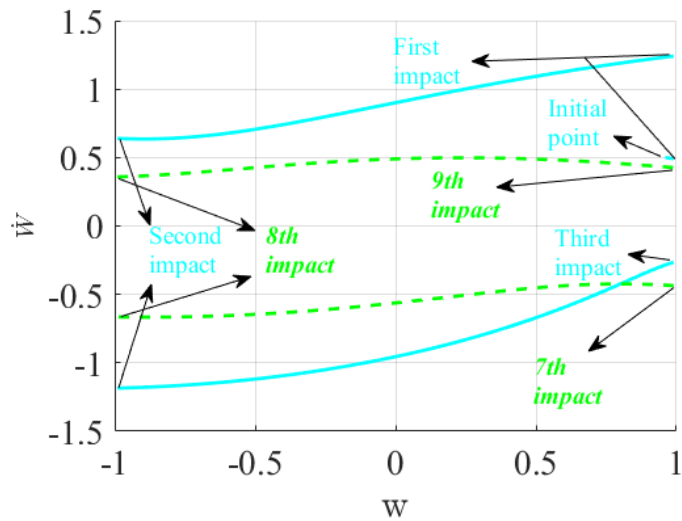

Fig. 6 The phase trajectory over 30 seconds

## 4. Efficiency metric

The efficiency of the VI-NES is measured by the percentage of excitation energy it absorbs and dissipates over an extended period. The relative energy is defined as the ratio between the instantaneous energy and the initial energy:

$$E_r = \frac{x_1^2 + \dot{x}_1^2 + \varepsilon \dot{x}_2^2 + \varepsilon \lambda \int_0^t \dot{x}_1^2 dt + c_e \int_0^t \dot{w}_1^2 dt}{\dot{x}_{1,0}^2} \tag{13}$$

A commonly used metric is the ratio of the energy dissipated by the VI-NES to the total energy of the system.

$$\bar{E}_{VI-NES} = \frac{\int_0^\infty E_r}{t} \tag{14}$$

The initial excitation can significantly impact the effectiveness of the VI-NES. Fig. 4 illustrates how the absorber's efficiency depends on the initial excitation. Fig. 2 shows the energy ratio after 30 seconds for different values of $\dot{x}_{1,0}$, highlighting the high sensitivity of the energy ratio to variations in the initial excitation.

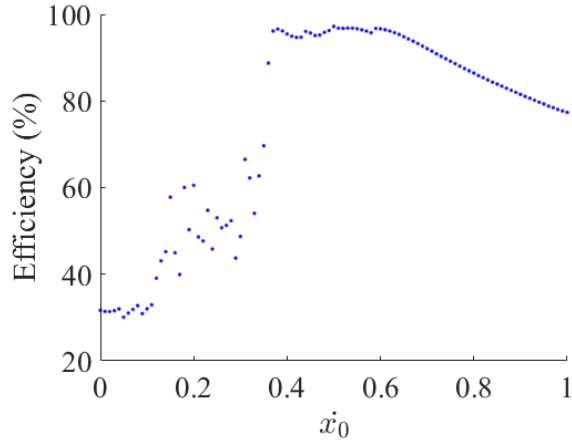

Fig. 7  $E_r$ values for different $\dot{x}_{1,0}$ with $\lambda$ =0.2, $c_e$ =0.05, and $\varepsilon$ =0.05.

## 5. Optimization formulation

The efficiency of an absorber is greatly influenced by the system's initial conditions. While deterministic optimization works well for a specific initial condition, it does not guarantee optimal performance under varying conditions. Thus, an alternative method is needed to handle uncertain initial conditions. To address this, the following multi-objective optimization approach is proposed:

$$\max E\left(\bar{E}_{VI-NES}(X_D.X_A)\right), \min \sigma \qquad (15)$$
$$\text{s.t.} \mu_{min}^D \leq \mu^D \leq \mu_{max}^D$$

Here, E represents the expected value, while $X_D$ and $X_A$ are random design parameters, such as the return coefficient, and random variables, such as initial conditions, respectively. The $\mu_D$ values denote the means of the random design parameters $X_D$, which are optimized to maximize the expected value and minimize the standard deviation $\sigma$. For clarity, the main steps of the optimization algorithm are illustrated in Fig. 3. This algorithm is based on a non-dominated sorting genetic algorithm that enhances the adaptive fitness of the population. By repeatedly using random aleatory and design variables, a Monte Carlo simulation is conducted to determine the probability of different efficiencies. This technique allows for assessing the impact of uncertainty on the performance of the NES. The expected value and standard deviation are calculated using the following equation:

$$E(\bar{E}_{\text{VI-NES}}(X_D.X_A)) \approx \frac{\sum_{i=1}^{N_{\text{MC}}} E(P_i)}{N_{\text{MC}}}$$

$$\sigma = \sqrt{\frac{\sum_{i=1}^{N_{\text{MC}}} (E(P_i) - \bar{E}_{\text{VI-NES}}(X_D.X_A))^2}{N_{\text{MC}} - 1}} \qquad (16)$$

where $P_i$ represents the $N_{\text{MC}}$ Monte Carlo samples centered around $\mu_D$

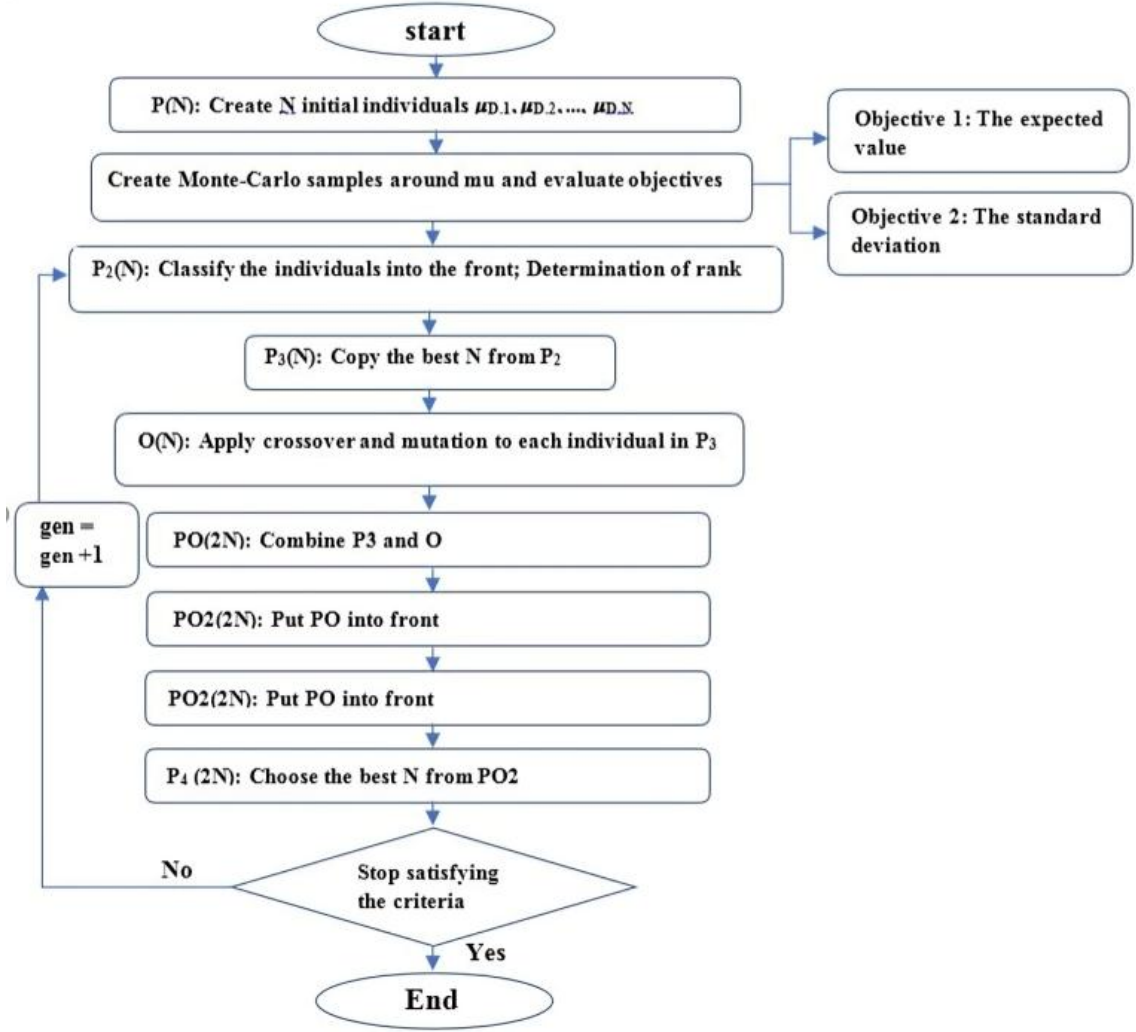

Fig. 8 Stochastic optimization algorithm

## 6. Formulations

$\dot{x}_{1,0}$ is treated as an aleatory variable, the problem is formulated as follows:

$$maxE(\overline{E}_{\text{VI-NES}}(\kappa, L_c, c_e, \dot{x}_{1,0})) \quad (17)$$
$$\text{s.t.} \ 0.001 \leq \mu_\kappa \leq 1$$
$$0.001 \leq \mu_{L_c} \leq 1$$
$$0.001 \leq \mu_{c_e} \leq 1$$

The design variables $\kappa$ and $L_c$ follow normal distributions: $\kappa \sim N\ (\mu_k, (2.97\times10^{-3})^2)$, $L_c \sim N\ (L_c, (2.97\times10^{-3})^2)$, and $c_e \sim N\ (c_e, (2.97\times10^{-3})^2)$. The standard deviations are set to 3% of the range of each variable. The initial velocity of the LO follows a uniform distribution: $(x_{1,0} \sim U\ (0.1, 1))$.

All parameters are set to c=0.2 and ω= 1. The expected value E ($E_{VI\text{-}NES}$) is determined using 1000 Monte Carlo samples. Although $\varepsilon$ can be a design parameter, previous studies [25, 26] have shown that greater mass improves performance and efficiency. Therefore, $\varepsilon$ is set to 0.05 in this study.

## 7. Comparison with deterministic optimality

The optimization results are summarized in Table 1. The energy ratio ($E_r$) is shown after 15 seconds for different values of ($\kappa$) and ($L_c$) at three distinct initial velocities. Fig. 9 illustrates the efficiency of the VI-NES for various initial velocities. The $c_e$ is taken as 0.05 for this figure. Fig. 10a illustrates the $E_r$ values for varying ce and kappa levels after 15 seconds. The data indicates that lower $c_e$ values result in higher efficiency compared to higher ce values. To further validate these findings, $E_r$ was also calculated for different $c_e$ and $L_c$ values in Fig. 10b.

Table 1: Comparison of Optimal Solutions for VI-NES under Stochastic and Deterministic Conditions

| Parameters | Stochastic | Deterministic | | |
| --- | --- | --- | --- | --- |
| | | $\dot{x}_{1.0} = 0.1$ | $\dot{x}_{1.0} = 0.55$ | $\dot{x}_{1.0} = 1$ |
| $\kappa$ | 0.39 | 0.54 | 0.43 | 0.25 |
| $L_c$ | 0.68 | 0.27 | 0.98 | 1 |
| $c_e$ | 0.013 | 0.06 | 0.013 | 0.011 |
| E($E_{VI\text{-}NES}$)(%) | 71.04 | 59.66 | 69.74 | 64.56 |
| 95% CI min | 70.43 | 58.96 | 68.77 | 63.56 |
| 95% CI max | 71.64 | 60.36 | 70.71 | 65.57 |
| $\sigma(E_{VI\text{-}NES})$ (%) | 9.28 | 11.22 | 15.59 | 16.17 |

The results are derived from both stochastic and deterministic optimization methods at three different initial velocities (see Fig. 5). The initial positions of the linear oscillator and the impact element are $x_{1,0}=0$ and $x_{2,0}=0.97$, respectively.

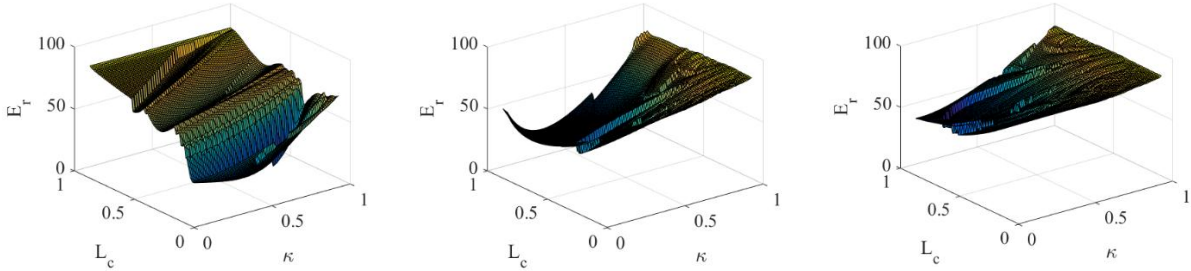

Fig. 9 The efficiency of the VI-NES for (a) $\dot{x}_{1.0}=0.1$, (b) $\dot{x}_{1.0}=0.55$, (c) $\dot{x}_{1.0}=1$

To visually demonstrate the performance of the VI-NES system with optimal parameter values, histograms, and cumulative distribution function (CDF) plots were created. These plots graphically represent how effectively the VI-NES absorbs and dissipates energy from the vibrating system (see Fig. 6).

The histograms display the distribution of energy levels before and after the VI-NES activation, highlighting the significant energy reduction achieved by the device. The CDF plots further illustrate the cumulative distribution of energy levels over time, emphasizing the VI-NES's sustained energy dissipation capabilities.

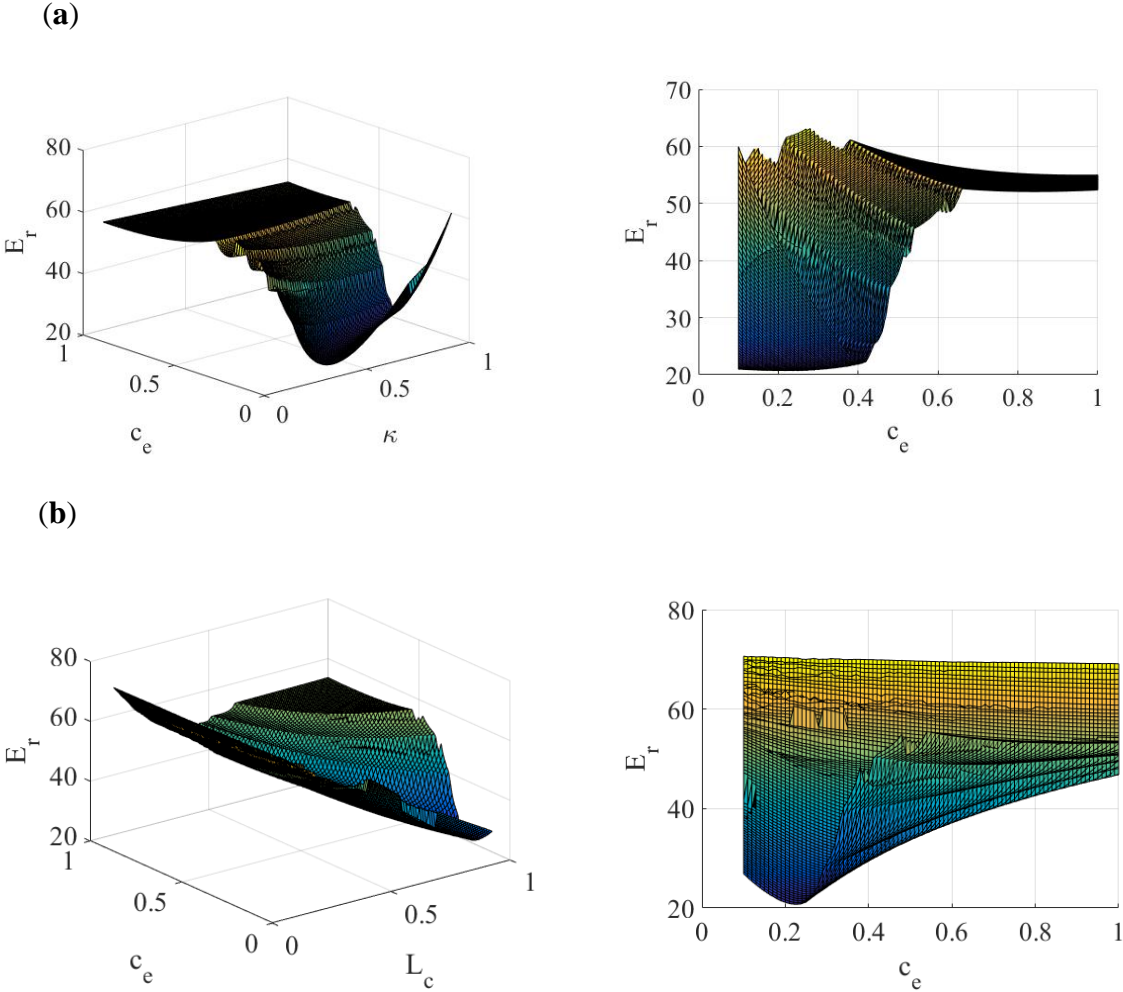

Fig. 10 The system's energy ratio for different values of (a) $\kappa$ and $c_e$, $L_c=1$ and (b) $L_c$ and $c_e$, $\kappa=0.6$.

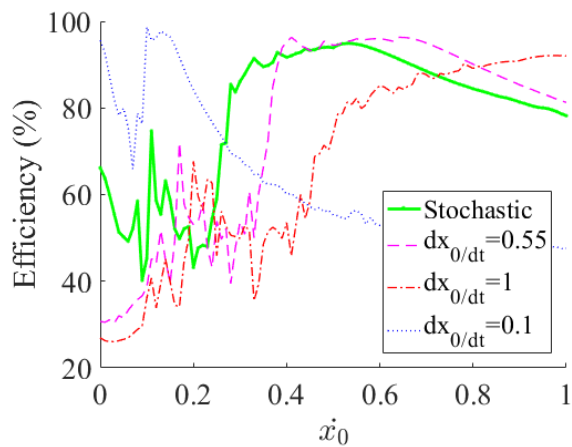

Fig. 11 Evaluating the efficiency of an absorber created using stochastic optimization compared to one developed using deterministic optimization.

In stochastic optimal design, the objective is to adjust design parameters to ensure significant Total Energy Transfer (TET) across the entire parameter space. This contrasts with the deterministic method, which aims to achieve maximum efficiency at a specific point, resulting in higher performance near that point. To demonstrate the effect of initial velocity on VI-NES performance, a Monte Carlo simulation was conducted, evaluating NES performance for different random variables after 20 seconds (see Fig. 7). The samples were divided into three clusters, and the results indicate that the optimal stochastic design can achieve high efficiency in most scenarios.

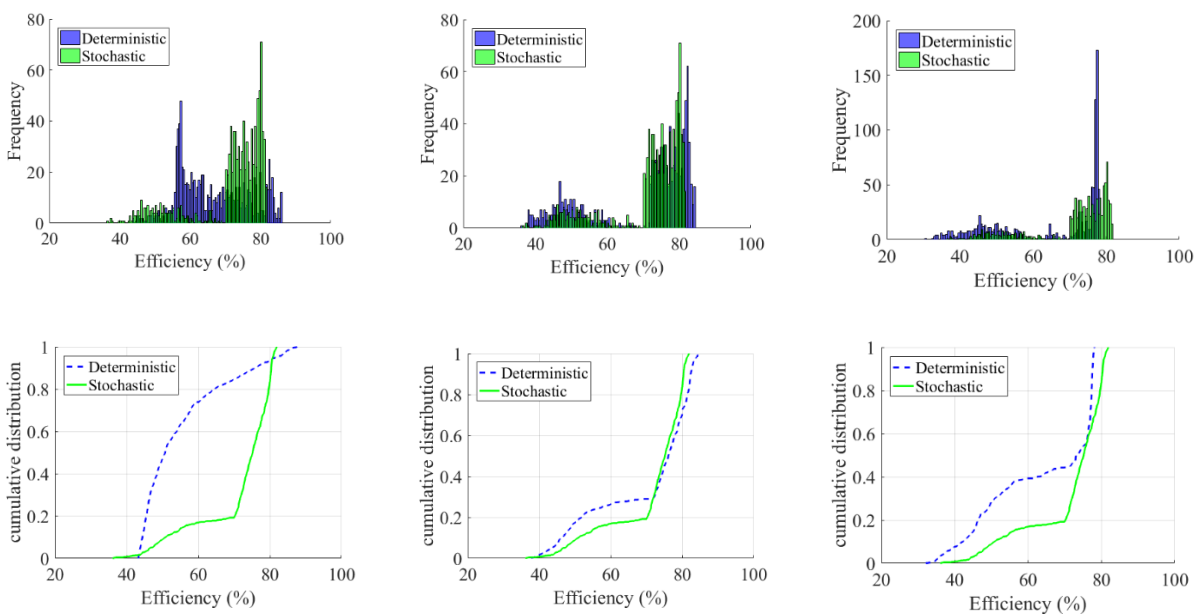

Fig. 12 Top: Histograms illustrating the efficiency of distributions near the stochastic and deterministic optima Bottom: The corresponding cumulative distributions for variables $\dot{x}_{1,0}= 0.1$ (left), $\dot{x}_{1,0}= 0.55$ (center), and $\dot{x}_{1,0} = 1$ (right).

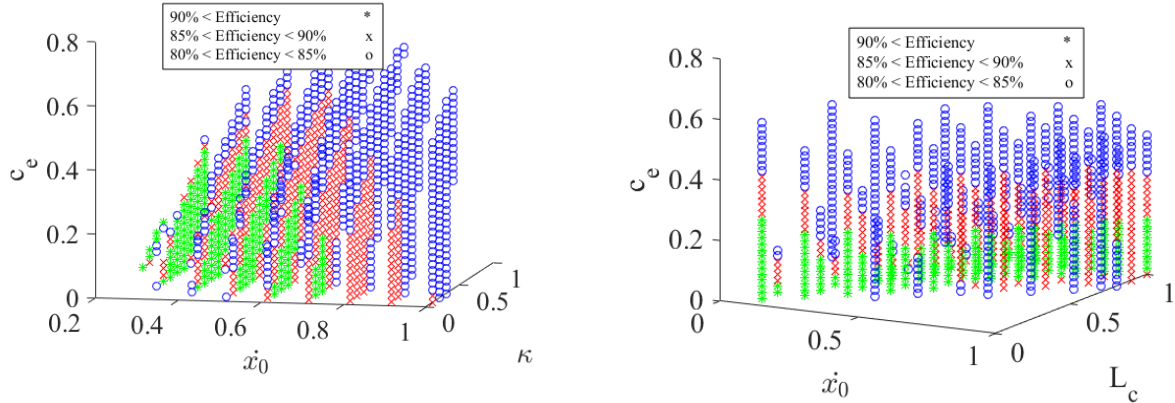

Fig. 13 Monte Carlo simulation to analyze the random variation of initial velocities considering different cavity lengths, coil parameters, and restitution coefficients over a period of 30 seconds.

## 8. Validation

To validate our methodology, we referred to a study by Qiu et al. [30], which is similar to our current work. Their research established criteria for determining the optimal cavity length in VI-NES systems, finding it to be 50 mm for the transient response. In our study, using a Genetic Algorithm, we determined the optimal cavity length to be 57.6 mm, which is close to their value. Fig 15 compares the results of our Genetic Algorithm with their criteria.

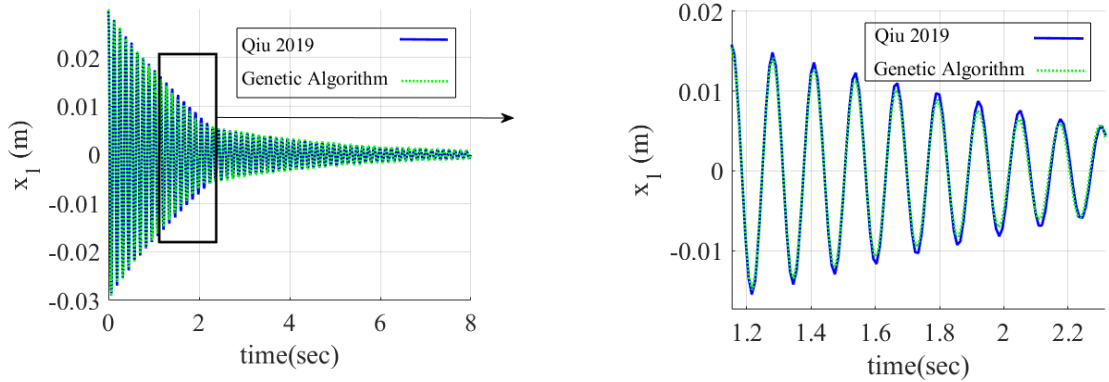

Fig. 14 Comparison of genetic algorithm and [30]

Since the goal of optimization is to minimize the mean system energy, we compared the energy ratios from both studies in Fig 14. In Qiu et al.'s study, the cavity length was the only design parameter considered. If both the cavity length ($L_c$) and the coefficient of restitution are optimized, their values would be 54.7 mm and 0.586, respectively. These results are illustrated in Fig 15.

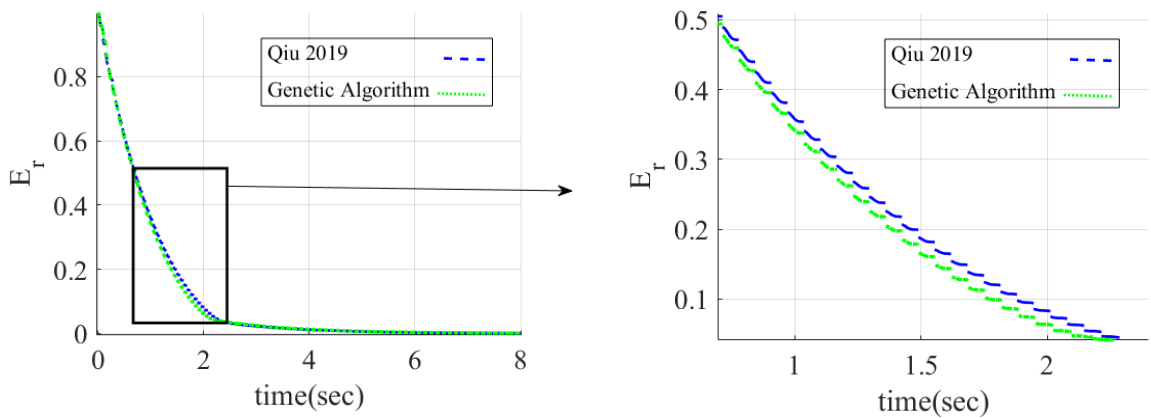

Fig. 15 The energy ratio of the system is optimized using an approximation method (dotted line) [30] and genetic algorithm (dotted line).

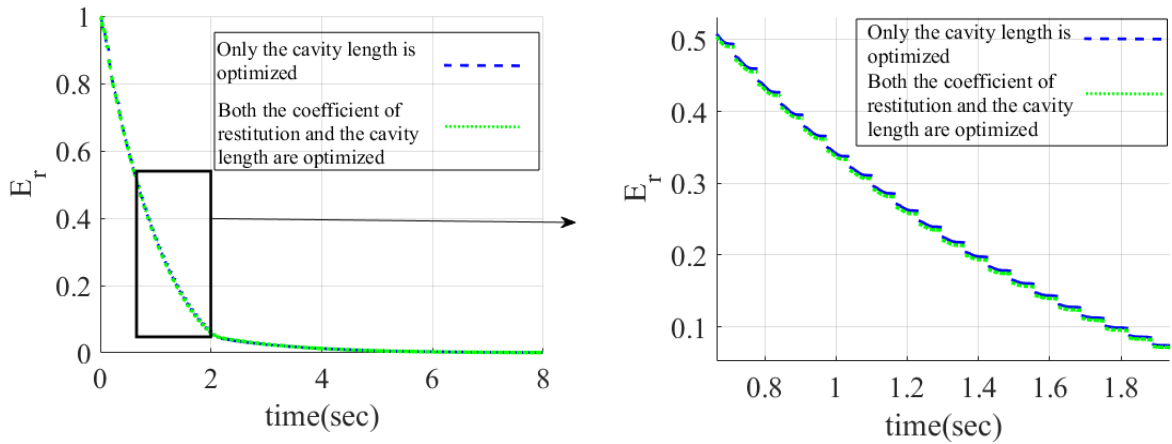

Fig. 16 The impact of optimizing the coefficient of restitution.

Although the primary objective of the optimization is to reduce the system's energy, it effectively minimizes the amplitude of the linear oscillator. As shown in Fig. 17, the amplitude response is minimized when both parameters (restitution coefficient and cavity length) are optimized using a Genetic Algorithm (GA). When only the cavity length is optimized, the GA method performs better than the approximate approach introduced in [30].

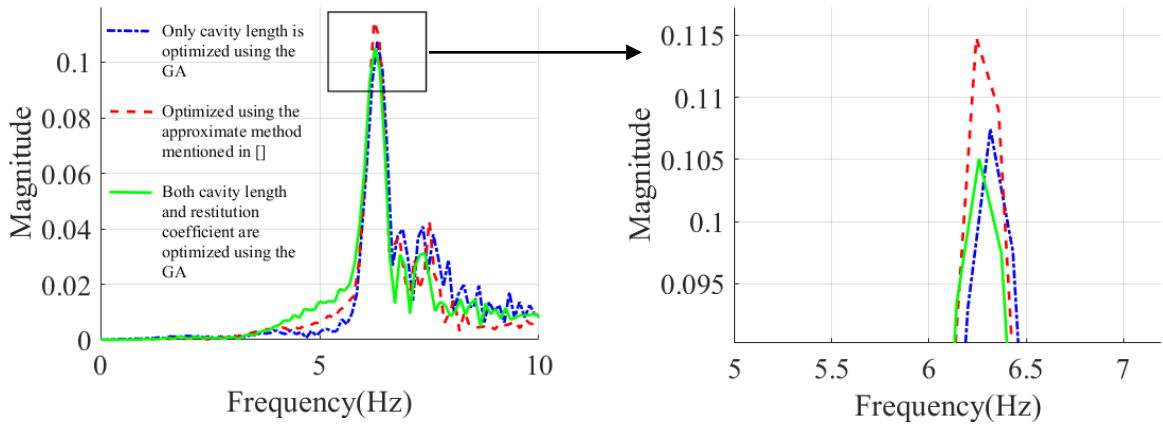

Fig. 17 Comparison of two optimization methods using frequency response

As previously noted, the optimal systems exhibit a specific dynamic behavior characterized by two impacts per cycle. Fig. 18a and 18b illustrate this behavior through relative displacement and wavelet transform, respectively.

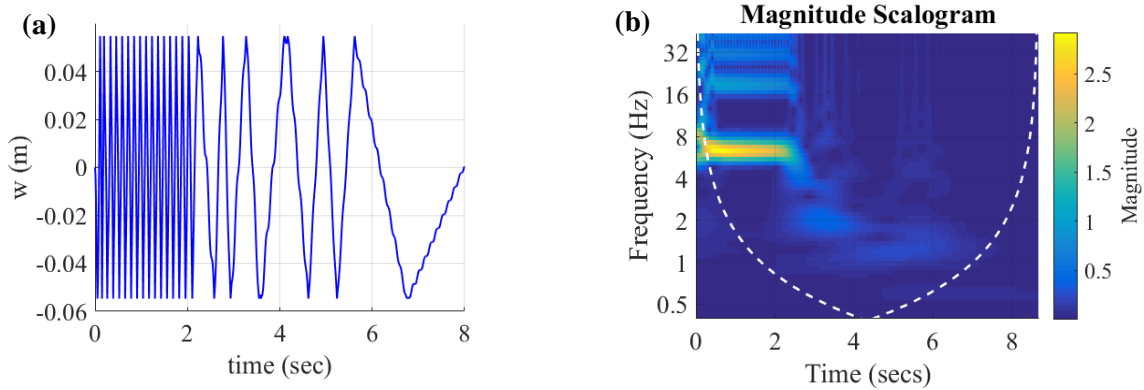

Fig. 18 The optimized dynamic behavior of the system using GA: (a) relative displacement, and (b) wavelet transform of the relative displacement.

## 9. Conclusion

This study has demonstrated the effectiveness of optimizing the coil parameter ($c_e$) the restitution coefficient and cavity length using a Genetic Algorithm (GA) to minimize the amplitude response of a linear oscillator equipped with a Vibro-Impact Nonlinear Energy Sink (VI-NES) and a coil. The results indicate that the GA method outperforms approximate approaches, particularly when both parameters are optimized simultaneously. The dynamic behavior of the optimized systems, characterized by two impacts per cycle, was thoroughly analyzed using relative displacement and wavelet transform techniques. The study also underscored the crucial influence of initial conditions on the VI-NES's efficiency, stressing the importance of robust optimization techniques to manage uncertainties. The effect of the coil on energy dissipation is examined by considering the coil parameter ($c_e$). It is shown that high values of ce decrease the efficiency of the absorber. The coil functions similarly to a damper attached to the magnet; thus, elevated values of $c_e$ reduce the velocity of the magnet, which in turn diminishes the efficiency of the vibration isolating. Validation against existing studies confirmed the reliability of the proposed optimization method.

This research offers valuable insights into designing and optimizing VI-NES systems for effective vibration mitigation.

**Reference**

[1]  Li H, Li A, Kong X, Xiong H. Dynamics of an electromagnetic vibro-impact nonlinear energy sink, applications in energy harvesting and vibration absorption. Nonlinear Dyn. 2022;108:1027-43.

[2]  Li H, Li S, Ding Q, Xiong H, Kong X. An electromagnetic vibro-impact nonlinear energy sink for simultaneous vibration suppression and energy harvesting in vortex-induced vibrations. Nonlinear Dyn. 2024;112:5919-36.

[3]  Wang Z, Chai X, Peng S, Wang B, Zhang L. A novel tuned liquid mass damper for low-frequency vertical vibration control: Model experiments and field tests. Mech Syst Signal Process. 2024 Nov;220:111702.

[4]  Li Z, Ma R, Xu K, Han Q. Closed-form solutions for the optimal design of lever-arm tuned mass damper inerter (LTMDI). Mech Syst Signal Process. 2024 Jan;206:110889.

[5]  Manzoni S, Berardengo M, Boccuto F, Vanali M. Piezoelectric-shunt-based approach for multi-mode adaptive tuned mass dampers. Mech Syst Signal Process. 2023 Oct;200:110537.

[6]  Rezazadeh H, Jafarzadeh V, Atabakhsh S, Dogani Aghcheghloo P. A novel passive nonlinear two-DOF internal resonance-based tuned mass damper. Mech Syst Signal Process. 2023 Dec;204:110788.

[7]	Zhang J, Xie F, Ma Z, Liu XJ, Zhao H. Design of parallel multiple tuned mass dampers for the vibration suppression of a parallel machining robot. Mech Syst Signal Process. 2023 Oct;200:110506.

[8]	Long Z, Shen W, Zhu H. On energy dissipation or harvesting of tuned viscous mass dampers for SDOF structures under seismic excitations. Mech Syst Signal Process. 2023 Apr 15;189:110087.

[9]	Den Hartog J. Mechanical vibrations. New York: McGraw-Hill; 1940.

[10]	Liu Q, Liu H, Zhang J. Optimization of nonlinear energy sink using Euler-buckled beams combined with piezoelectric energy harvester. Mech Syst Signal Process. 2025 Jan 15;223:111812.

[11]	Li H, Li S, Zhang Z, Xiong H, Ding Q. Effectiveness of vibro-impact nonlinear energy sinks for vibration suppression of beams under traveling loads. Mech Syst Signal Process. 2025 Jan 15;223:111861.

[12]	Dekemele K, Giraud-Audine C, Thomas O. A piezoelectric nonlinear energy sink shunt for vibration damping. Mech Syst Signal Process. 2024 Nov 1;220:111615.

[13]	Guo M, Tang L, Mace B, Inman DJ. Vibration suppression performance of parallel magnetic nonlinear energy sinks under impulse excitations. Mech Syst Signal Process. 2025 Jan 1;222:111810.

[14]	Zeng YC, Ding H, Ji JC, Chen LQ. Theoretical and experimental study of a stable state adjustable nonlinear energy sink. Mech Syst Signal Process. 2024 Jul 1;216:111470.

[15]	Geng X, Ding H, Jing X, Mao X, Wei K, Chen L. Dynamic design of a magnetic-enhanced nonlinear energy sink. Mech Syst Signal Process. 2023 Feb 15;185:109813.

[16] Vakakis AF, Gendelman OV, Bergman LA, McFarland DM, Kerschen GG, Lee YS. Nonlinear targeted energy transfer in mechanical and structural systems. In: Solid mechanics and its applications. Vol. 156. Dordrecht, Netherlands: Springer; 2009.

[17] Yegorov I, Uden A, Yurchenko D. Optimal performance comparison of nonlinear energy sinks and linear tuned mass dampers. In: Proceedings of the ASME 2019 International Design Engineering Technical Conferences and Computers and Information in Engineering Conference: 17th International Conference on Multibody Systems, Nonlinear Dynamics, and Control; August 18-21, 2019; Anaheim, California, USA. New York: ASME; 2019. p. V009T12A003

[18] Gourc E, Michon G, Seguy S, Berlioz A. Experimental investigation and design optimization of targeted energy transfer under periodic forcing. J Vibr Acoust. 2014;136(3):031009.

[19] Li T, Seguy S, Berlioz A. Dynamics of cubic and vibro-impact nonlinear energy sink: Analytical, numerical, and experimental analysis. J Vibr Acoust. 2016;138(1):011008.

[20] Gourc E, Michon G, Seguy S, Berlioz A. Targeted energy transfer under harmonic forcing with a vibro-impact nonlinear energy sink: Analytical and experimental developments. J Vibr Acoust. 2015;137(5):051007.

[21] Li T, Seguy S, Berlioz A. Optimization mechanism of targeted energy transfer with vibro-impact energy sink under periodic and transient excitation. Nonlinear Dynam. 2016;87(2):2415-33.

[22] Afsharfard A. Application of nonlinear magnetic vibro-impact vibration suppressor and energy harvester. Mech Syst Signal Process. 2018;98:371-81.

[23]     Wu M, Zhang J, Wu H. Vibration mitigation and energy harvesting of vibro–impact dielectric elastomer oscillators. Int J Mech Sci. 2024;265:108906.

[24]     Li H, Li A. Potential of a vibro-impact nonlinear energy sink for energy harvesting. Mech Syst Signal Process. 2021;159:107827. doi:10.1016/j.ymssp.2021.107827.

[25]     E. Boroson, S. Missoum, Stochastic optimization of nonlinear energy sinks, Struct. Multidiscip. Optim. 55 (2) (2017) 633–646.

[26]     B. Pidaparthi, S. Missoum, 2018. Optimization of Nonlinear Energy Sinks for the Mitigation of Limit Cycle Oscillations, In 2018 Multidisciplinary Analysis and Optimization Conference (p. 3569).

[27]     Qian J, Chen L. Optimization for vibro-impact nonlinear energy sink under random excitation. Theor Appl Mech Lett. 2022 Sep;12(5):100364.

[28]     Wu P, Xiao J, Zhao Y. Power Spectrum Analysis and Optimization Design of Nonlinear Energy Sink Under Random Excitation. J Vib Eng Technol. 2023 Dec 11;12:5663-5673.

[29]     Bao FY, Cong CL. Suppressing random response of structure via inertial nonlinear energy sink. Eur Phys J Spec Top. 2024 Jun 13.

[30]     Qiu D, Seguy S, Paredes M. Design criteria for optimally tuned vibro-impact nonlinear energy sink. J Sound Vib. 2019 Mar 3;442:497-513.